\documentclass[12pt]{amsart}
\usepackage{enumerate}
\usepackage{amsmath}
\usepackage{amssymb}
\usepackage{amsfonts}
\usepackage{amsthm, upref}
\usepackage{graphicx}
\usepackage{url}

    \numberwithin{equation}{section}

\DeclareMathOperator{\sgn}{sgn}

\newcommand{\pp}{\mathbb{P}}
\newcommand{\ev}{\mathbb E}  

\theoremstyle{plain}
\newtheorem{thm}{Theorem}
\newtheorem{lemma}[thm]{Lemma}
\newtheorem{prop}[thm]{Proposition}
\newtheorem{cor}[thm]{Corollary}
\newtheorem{exm}[thm]{Example}

\theoremstyle{definition}

\theoremstyle{remark}

\renewcommand{\Im}{\operatorname{Im}}
\renewcommand{\Re}{\operatorname{Re}}

\begin{document}

\title[Implied Volatility Slope for L\'evy Models]{Small-maturity asymptotics for the
at-the-money implied volatility slope in L\'evy models}

\author[S.~Gerhold]{Stefan Gerhold}
\address{TU Wien, Wiedner Hauptstra\ss{}e 8--10/E105-1,
A-1040 Vienna, Austria}
\email{sgerhold@fam.tuwien.ac.at}
\author[I.~C.\ G\"ul\"um]{I.~Cetin G\"ul\"um}
\address{TU Wien, Wiedner Hauptstra\ss{}e 8--10/E105-1,
A-1040 Vienna, Austria}
\email{ismail.gueluem@tuwien.ac.at}
\author[A.~Pinter]{Arpad Pinter}
\address{TU Wien, Wiedner Hauptstra\ss{}e 8--10/E105-1,
A-1040 Vienna, Austria}
\email{arpad.pinter@tuwien.ac.at}

\thanks{We thank Jos\'e Fajardo, Peter Friz, Friedrich Hubalek, Andreas Kyprianou, and Mykhaylo Shkolnikov for helpful discussions,
 and gratefully acknowledge financial support from the
Austrian Science Fund (FWF) under grant P~24880-N25. Special
thanks are due to the anonymous referees for their comprehensive
and very helpful remarks, some of which led us to alter the core of the paper.}

\date{\today}

\begin{abstract}
  We consider the at-the-money strike derivative of implied volatility
  as the maturity tends to zero. Our main results quantify the behavior
  of the slope for infinite activity exponential L\'evy models including
  a Brownian component.
  As auxiliary results, we obtain asymptotic expansions of short maturity
  at-the-money digital call options, using Mellin transform asymptotics.
  Finally, we discuss when the at-the-money slope is consistent
  with the steepness of the smile wings, as given by Lee's moment formula.
\end{abstract}



\maketitle

\section{Introduction}

Recent years have seen an explosion of the literature on asymptotics
of option prices and implied volatilities (see, e.g.,~\cite{AnLi13,
FrGeGuSt11} for many references).
Such results are of practical relevance for fast model calibration,
qualitative model assessment, and parametrization design.
The small-time behavior of the \emph{level} of implied volatility in L\'evy models
(and generalizations)
has been investigated in great detail~\cite{BoLe02,FiFo12,FiGoHo12,FiGoHo14,Ro08,Ta11}.
We, on the other hand, focus on the at-the-money \emph{slope}
of implied volatility, i.e., the strike derivative, and investigate
its behavior as maturity becomes small.
For diffusion models, there typically exists a limiting smile
as the maturity tends to zero, and the limit slope is just the slope
of this limit smile (e.g., for the Heston model, this follows
from~\cite[Section~5]{Du10}). Our focus is, however, on exponential L\'evy models.
There is no limit smile here that one could differentiate, as
the implied volatility blows up off-the-money~\cite{Ta11}.
In fact, this is a desirable
feature, since in this way L\'evy models are better suited to capture
the steep short maturity smiles observed in the market.
But it also implies that the limiting slope cannot be deduced directly
from the behavior of implied volatility itself, and requires
a separate analysis. (Note that a limiting smile does exist
if maturity and log-moneyness tend to zero jointly in an appropriate
way~\cite{MiTa12}.)

It turns out that the presence of a Brownian component has a decisive
influence: Without it,
the ATM (at-the-money) slope explodes (under mild conditions).
The blowup is of order~$T^{-1/2}$ for many models,
but may also be slower (CGMY model with $Y\in(1,2)$, e.g.; see Example~\ref{ex:cgmy}).
Our main results are on L\'evy models
\emph{with} a Brownian component, though. We provide a result (Corollary~\ref{cor:main}
in Section~\ref{se:levy mellin}) that translates the asymptotic behavior of
the moment generating function to that of the ATM slope. When applied
to concrete models, we see that the slope may converge to a finite limit
(Normal Inverse Gaussian, Meixner, CGMY models), or explode at a rate slower than $T^{-1/2}$
(generalized tempered stable model; this kind of behavior
seems to be the most realistic one, see~\cite{BaFrGa16}).
Note that several studies~\cite{Ai02,AiJa10,CaWu03}
highlight the importance of a Brownian component when fitting to historical data
or option prices. In particular, in many pure jump L\'evy models ATM implied
volatility converges to zero as $T\downarrow0$ (see Proposition~5 in~\cite{Ta11}
for a precise statement), which seems undesirable.

From a practical point of view, the asymptotic slope is a useful ingredient
for model calibration: E.g., if the market slope is negative, then
a simple constraint on the model parameters forces the (asymptotic) model slope
to be negative, too.
Our numerical tests show that the sign of the slope is reliably
identified by a first order asymptotic approximation, even if the maturity
is not short at all.
With our formulas, the asymptotic slope (and, of course, its sign)
can be easily determined from the model parameters. For instance,
the slope of the NIG (Normal Inverse Gaussian) model is positive 
if and only if the skewness parameter satisfies $\beta>-\tfrac12$.

To obtain these results, we investigate the asymptotics of at-the-money
digital calls; their relation to the implied volatility slope is well known.
While, for L\'evy processes
$X$, the small-time behavior of the transition probabilities $\pp[X_T\geq x]$
(in finance terms, digital call prices)
has been well studied for $x \neq X_0$ (see, e.g.,~\cite{FiHo09} and
the references therein),
not so much is known for $x=X_0$.
Still, first order asymptotics of $\pp[X_T\geq X_0]$ are available, and this suffices
if there is no Brownian component.
If the L\'evy process has a Brownian component, then it is well known that
$\lim_{T\to0}\pp[X_T\geq X_0]=\tfrac12$.
In this case, it turns out that the second order term of $\pp[X_T\geq X_0]$ is required
to obtain slope asymptotics.
For this, we use a novel approach involving the Mellin transform (w.r.t.\ time)
of the transition probability (Sections~\ref{se:mellin} and~\ref{se:levy mellin}).
We believe that this method is of wide applicability to other problems
involving time asymptotics of L\'evy processes, and hope to elaborate
on it in future work.

Finally, we consider the question whether a positive at-the-money slope
requires the right smile wing to be the steeper one, and vice versa.
Wing steepness refers to large-strike asymptotics here.
It turns out that this is indeed the case for several of the infinite activity models we
consider. This results in a qualitative limitation
on the smile shape that these models can produce.

One of the few other works dealing with small-time
L\'evy slope asymptotics is the comprehensive recent paper
by Andersen and Lipton~\cite{AnLi13}. Besides many other problems on various models
and asymptotic regimes, they study the small-maturity
ATM digital price and volatility slope for the tempered stable model
(Propositions~8.4 and~8.5 in~\cite{AnLi13}). This
includes the CGMY model as a special case (see Example~\ref{ex:cgmy} for details).
Their proof method is entirely different from ours,
exploiting the explicit form of the characteristic
function of the tempered stable model. Using mainly the dominated convergence
theorem, they also analyze the convexity. We, on the other hand, assume a certain
asymptotic behavior of the characteristic function, and use its explicit expression
only when calculating concrete examples.
Our approach covers, e.g., the ATM slope of
the generalized tempered stable, NIG, and Meixner models
without additional effort.

The recent preprint~\cite{FiOl15} is also closely related to our work. There, the Brownian
component is generalized to stochastic volatility. On the other hand, the assumptions
on the L\'evy measure exclude, e.g., the NIG and Meixner models.
Section~\ref{se:ex} has additional comments on how our results compare to those
of~\cite{AnLi13} and~\cite{FiOl15}.
Al{\`o}s et al.~\cite{AlLeVi07} also
study the small time implied volatility slope under stochastic volatility and jumps, but the latter are
assumed to have finite activity, which is not our focus.
Results on the \emph{large} time slope can be found in~\cite{FoJaFi11};
see also~\cite{Ga06}, p.~63f.

\section{Digital call prices}\label{se:dig}

We denote the underlying by~$S=e^X$, normalized to $S_0=1$,
and the pricing measure by~$\pp$.
W.l.o.g.\ the interest rate is set to zero, and so $S$
is a $\pp$-martingale. 
Suppose that the log-underlying
$X=(X_t)_{t\geq0}$ is a L\'evy process with characteristic triplet $(b,\sigma^2,\nu)$
and $X_0=0$.
The moment generating function (mgf) of~$X_T$ is
\begin{equation*}
  M(z,T) = \ev[e^{z X_T}] = \exp\left(T \psi(z) \right),
\end{equation*}
where
\begin{equation}\label{eq:psi}
  \psi(z)= \tfrac12 \sigma^2 z^2 +bz 
  +  \int_{\mathbb{R}}(e^{zx} - 1 - z x )\nu(dx).
\end{equation}
This representation is valid if the L\'evy process has a finite first moment, which
we of course assume, as even $S_t=e^{X_t}$ should be integrable.
If, in addition, $X$ has paths of finite variation, then $\int_{\mathbb{R}} |x| \nu(dx)<\infty$, and
\[
    \psi(z)= \tfrac12 \sigma^2 z^2 +b_0z 
  +  \int_{\mathbb{R}}(e^{zx} - 1 )\nu(dx),
\]
where the drift~$b_0$ is defined by
\[
  b_0 = b - \int_{\mathbb{R}} x\,\nu(dx).
\]
The following theorem collects some results about the small-time behavior
of $\pp[X_T \geq 0]$. All of them are known, or easily obtained from known results.
We are mainly interested in the case where $S=e^X$ is a martingale, and so
$\pp[X_T \geq 0]$ has the interpretation of an at-the-money digital call price.
Still, we mention
that this assumption is not necessary for parts (i)-(iv).
In part~(iv), the following condition from~\cite{RoTa11} is used:
\begin{align*}
  (\mathbf{H}\text{-}\alpha)& \text{ The L\'evy measure~$\nu$ has a density $g(x)/|x|^{1+\alpha}$, where~$g$
is a non-negative} \\
&\ \text{measurable function admitting left and right limits at zero:} \\
&\ c_+ := \lim_{x\downarrow0}g(x), \quad c_- := \lim_{x\uparrow0}g(x), \quad
 \text{with}\ \ c_+ + c_->0.
\end{align*}
\begin{thm}\label{thm:dig}
Let~$X$ be a L\'evy process with characteristic triplet $(b,\sigma^2,\nu)$
and $X_0=0$.
  \begin{itemize}
    \item [(i)] If~$X$ has finite variation, and $b_0\neq0$, then
      \[
       \lim_{T\downarrow0}\pp[X_T \geq 0]=
       \begin{cases}
         1, & b_0>0 \\
         0, & b_0<0.
       \end{cases}
      \]
      \item [(ii)] If $\sigma>0$, then $\lim_{T\downarrow0}\pp[X_T \geq 0]=\tfrac12$.
      \item [(iii)] If~$X$ is a L\'evy jump diffusion, i.e., it has finite activity jumps and $\sigma>0$,
      then
      \[
         \mathbb{P}[X_T \geq 0] = \frac12 + \frac{b_0}{\sigma \sqrt{2\pi}} \sqrt{T}
          + O(T), \quad T\downarrow 0.
      \]
      \item [(iv)] Suppose that $\sigma=0$ and that $(\mathbf{H}$-$\alpha)$ holds for
      some $\alpha \in [1,2)$. If $\alpha=1$, we additionally assume $c_-=c_+=:c$ and
      $\int_0^1 x^{-1}|g(x)-g(-x)|dx<\infty$. Then
      \[
        \lim_{T \downarrow 0} \mathbb{P}[X_T \geq 0]= 
          \begin{cases}  
	       \frac 12+ \frac {1} {\pi}\arctan \frac {b^*} {\pi c}  & \text{if} \ \ \alpha=1,  \\
        \frac 12 +\frac \alpha \pi \arctan \bigl(\beta \tan \bigl( \frac{\alpha \pi}{2} \bigr) \bigr)     &           \text{if} \ \ \alpha \neq 1,
        \end{cases}
      \]
      where $b^*=b-\int_0^\infty (g(x)-g(-x))/x\, dx$ and $\beta=(c_+-c_-)/(c_+ + c_-)$.
      \item [(v)] If $e^X$ is a martingale and
      the L\'evy measure satisfies $\nu(dx)=e^{-x/2}\nu_0(dx)$,
        where $\nu_0$ is a symmetric measure, then
        \[
          \mathbb{P}[X_T \geq 0]  = \Phi(-\sigma_{\mathrm{imp}}(1,T) \sqrt{T}/2),
        \]
        where $\Phi$ denotes the standard Gaussian cdf.
  \end{itemize}
\end{thm}
\begin{proof}
  (i) We have $\pp[X_T \geq 0]=\pp[T^{-1}X_T \geq 0]$, but $T^{-1}X_T$ converges
  a.s.\ to~$b_0$, by Theorem~43.20 in~\cite{Sa99}.
  
  (ii) If $\sigma>0$, then $T^{-1/2}X_T$ converges in distribution to a 
  centered Gaussian random variable
  with variance~$\sigma^2$ (see~\cite{Sa99}).
  For further CLT-type results in this vein, see~\cite{DoMa02,GeKlPoSh15}.
  
  (iii) Conditioning on the first jump time~$\tau$, which has an exponential distribution,
  we find
  \begin{align}
     \pp[X_T\geq0] &= \pp[X_T\geq0 | \tau\leq T]\cdot\pp[\tau \leq T]
     + \pp[X_T\geq0 | \tau > T]\cdot\pp[\tau > T] \notag \\
     &= O(T)+ \pp[\sigma W_T +b_0T  \geq0](1+O(T)) \notag \\
     &= \pp[\sigma W_T +b_0T  \geq0] + O(T) \notag \\
     &= \Phi(b_0\sqrt{T}/\sigma) + O(T). \label{eq:cond}
  \end{align}
  Now apply the expansion
  \begin{equation}\label{eq:Phi}
    \Phi(x)=\frac 12 + \frac {x}{\sqrt{2\pi}}+O(x^3), \quad x \rightarrow 0.
  \end{equation}  
  (iv) By Proposition~1 in~\cite{RoTa11},
  the rescaled process $X_t^{\varepsilon,\alpha}:= \varepsilon^{-1}X_{\varepsilon^\alpha t}$
  converges in law to a strictly $\alpha$-stable process $X_t^{*,\alpha}$ as $\varepsilon \downarrow  0$.
  Therefore
  \[
         \lim_{T \downarrow 0} \pp[X_T \geq 0] =\lim_{\varepsilon \downarrow 0} \pp[\varepsilon^ 
         {-1 }X_{\varepsilon^\alpha}  \geq 0]= \pp[X^{*,\alpha}_1 \geq 0],
   \]
   and it suffices to evaluate the latter probability.
   For $\alpha=1$, $X^{*,1}_1$ has a Cauchy distribution with
   characteristic exponent
   \[
     \log \ev[\exp(iuX^{*,1}_1)] = ib^*u - \pi c |u|,
   \]
   hence $\pp[X^{*,1}_1 \geq 0]=\frac{1}{\pi} \arctan \tfrac{b^*}{\pi c}$.
   (Our $b^*$ is denoted $\gamma^*$ in~\cite{RoTa11}.)
   
   If $1<\alpha<2$, then $X^{*,\alpha}_1$ has a strictly stable distribution with
   characteristic exponent
   \[
     \log \ev[\exp(iuX^{*,\alpha}_1)] =
     -|du|^{\alpha} \Bigl(1-i \beta \mathop{\rm sgn}(u)\tan\bigl( \frac{\alpha \pi}{2} \bigr) \Bigr),
   \]
   where
   \[
     d_{\pm}^\alpha= -\Gamma(-\alpha)\cos\bigl( \frac{\alpha \pi}{2} \bigr) c_\pm \geq 0, \quad d^{\alpha}= d_+^\alpha + d_-^\alpha,  \quad \beta= \frac{d_+^\alpha - d_-^\alpha}{d^\alpha} \in (-1,1).
   \]
   The desired expression for $\pp[X^{*,\alpha}_1 \geq 0]$ then follows from~\cite{DaIb71}.
   See~\cite{FiFo12} for further related references.
   
  (v) Under this assumption, the market model is symmetric in the sense
  of~\cite{Fa14,FaMo06}. The statement is Theorem~3.1 in~\cite{Fa14}.
\end{proof}

The variance gamma model and the CGMY model with $0<Y<1$ are examples of finite
variation models (of course, only when $\sigma=0$), and so part~(i)
of Theorem~\ref{thm:dig} is applicable. Part~(iii) is applicable, clearly,
to the well-known jump diffusion models by Merton and Kou. In Section~\ref{se:ex},
we will discuss two examples for part~(iv) (NIG and Meixner).

\section{Implied Volatility Slope and Digital Options with
Small Maturity}\label{se:trans}

The (Black-Scholes) implied volatility
is the volatility that makes the Black-Scholes call price
equal the call price with underlying~$S$:
\[
  C_{\mathrm{BS}}(K,T,\sigma_{\mathrm{imp}}(K,T))
    = C(K,T) := \ev[(S_T-K)^+].
\]
Since no explicit expression is known for $\sigma_{\mathrm{imp}}(K,T)$
(see~\cite{Ge13}), many authors have investigated approximations
(see, e.g., the references in the introduction).
The following relation between implied volatility slope and digital calls
is well known~\cite{Ga06}; we give a proof for completeness.
(Note that absolute continuity of~$S_T$ holds in all L\'evy models
of interest, see Theorem~27.4 in~\cite{Sa99}, and will be assumed throughout.)
\begin{lemma} \label{le:dig slope}
Suppose that the law of~$S_T$ is absolutely continuous for each $T>0$, and that
\begin{equation}\label{eq:ass2}
  \lim_{T\downarrow0}C(K,T)=(S_0-K)^+, \quad K>0.
\end{equation}
Then, for $T \downarrow 0$,
\begin{equation} \label{eq:dig to slope}
\partial_K \sigma_{\mathrm{imp}}(K,T)|_{K=1} \sim \sqrt{\frac{2\pi}{T}} \biggl( \frac 12- \mathbb{P}[S_T \geq 1]- \frac{\sigma_{\mathrm{imp}}(1,T) \sqrt{T}}{2\sqrt{2\pi}}+ O \bigl( \bigl(\sigma_{\mathrm{imp}}(1,T) \sqrt{T}\bigr)^2 \bigr) \biggr).
\end{equation}
\end{lemma}
\begin{proof}
By the implicit function theorem, the implied volatility slope has the representation
\[
\partial_K \sigma_{\mathrm{imp}}(K,T) = \frac{\partial_K C(K,T)-\partial_K C_{\mathrm{BS}}(K,T,\sigma_{\mathrm{\mathrm{imp}}}(K,T))}{\partial_\sigma C_{\mathrm{BS}}(K,T,\sigma_{\mathrm{imp}}(K,T))}.
\]
Since the law of~$S_T$ is absolutely continuous, the call price $C(K,T)$ is continuously differentiable
w.r.t.~$K$, and $\partial_K C(K,T) =-\mathbb{P}[S_T \geq K]$.
Inserting the explicit formulas for the Black-Scholes Vega and digital price,
and specializing to the ATM case $K=S_0=1$, we get
\[
\partial_K \sigma_{\mathrm{imp}}(K,T)|_{K=1} = \frac{\Phi(-\sigma_{\mathrm{imp}}(1,T)\sqrt{T}/2)-\mathbb{P}[S_T \geq 1]}{\sqrt{T}\varphi(\sigma_{\mathrm{imp}}(1,T)\sqrt{T}/2)},
\]
where $\Phi$ and $\varphi$ denote the standard Gaussian cdf and density, respectively.
By Proposition~4.1 in~\cite{RoRu09}, our assumption~\eqref{eq:ass2} implies
that the annualized implied volatility $\sigma_{\mathrm{imp}}(1, T) \sqrt T$
tends to zero as $T \downarrow 0$.
(The second assumption used in~\cite{RoRu09} are the no-arbitrage bounds
$(S_0-K)^+ \leq C(K,T) \leq S_0$, for $K,T>0$, but these are satisfied here because
our call prices are generated by the martingale~$S$.)
Using the expansion~\eqref{eq:Phi}
and $\varphi(x)= \frac{1}{\sqrt{2\pi}}+O(x^2)$, we thus obtain~\eqref{eq:dig to slope}.
\end{proof}
The asymptotic
relation~\eqref{eq:dig to slope}  is, of course, consistent with the small-moneyness expansion presented in~\cite{DeVaCiBo12}, where
$\sqrt{2\pi/T}\left(\frac12- \pp[S_T\geq K]\right)$
appears as second order term (i.e., first derivative) of implied volatility.

Lemma~\ref{le:dig slope} shows that, in order to obtain first
order asymptotics for the at-the-money (ATM) slope, we need first
order asymptotics for the ATM digital call price $\pp[S_T\geq1]$.
(Recall that $S_0=1$.)
For models where $\lim_{T\downarrow0}\pp[S_T\geq1]=\tfrac12$, we need
the second order term of the digital call as well, and the first order term
of $\sigma_{\mathrm{imp}}(1,T)\sqrt{T}$.
The limiting value $1/2$ for the ATM digital call is typical for diffusion
models (see~\cite{GeKlPoSh15}), and L\'evy processes that contain a Brownian motion.
For infinite activity models without diffusion component, $\pp[S_T\geq1]$
may converge to $1/2$ as well (e.g., in the CGMY model with $Y\in(1,2)$), but
other limiting values are also possible. See the examples in Section~\ref{se:ex}.

{}From part~(i) of Theorem~\ref{thm:dig} and Lemma~\ref{le:dig slope} we can immediately conclude
the following result. Note that we assume throughout that~$X$ is such that
$S=e^X$ is a martingale with $S_0=1$.
\begin{prop}\label{prop:fin var}
  Suppose that the L\'evy process~$X$ has finite variation
  (and thus, necessarily, that $\sigma=0$), and that $b_0\neq0$. Then
  the ATM implied volatility slope satisfies
  \[
    \partial_K \sigma_{\mathrm{imp}}(K,T)|_{K=1} \sim
      -\sqrt{\pi/2} \sgn(b_0) \cdot T^{-1/2},
    \quad T\downarrow0.
\]
\end{prop}
Note that $T^{-1/2}$ is the fastest possible growth order for the slope, in any model
(see Lee~\cite{Le05}).

If~$X$ is a L\'evy jump diffusion with $\sigma>0$, then
by part (iii) of Theorem~\ref{thm:dig}, \eqref{eq:dig to slope}, and the fact that $\sigma_{\mathrm{imp}}\to\sigma$
(implied volatility converges to spot volatility),
we obtain the finite limit
\begin{equation}\label{eq:limit}
  \lim_{T\downarrow 0}\partial_K  \sigma_{\mathrm{imp}}(K,T)|_{K=1}=
  -\frac{b_0}{\sigma} - \frac{\sigma}{2}.
\end{equation}
(It is understood that the substitution $K=1$ is to be performed before the limit $T\downarrow 0$.)
Notice that the expression on the right hand side of~\eqref{eq:limit} does depend on the jump parameters,
because the drift~$b_0$, fixed by the condition $\ev[\exp(X_1)]=1$, depends on them.
Moreover, \eqref{eq:limit} is consistent with the formal calculation
of the variance slope
\[
  \lim_{T\downarrow 0}\partial_K \sigma_{\mathrm{imp}}^2(K,T)|_{K=1}=-2b_0-\sigma^2
\]
on p.~61f in~\cite{Ga06}.
In fact~\eqref{eq:limit} is well known for jump diffusions, see~\cite{AlLeVi07,Ya11}.

\section{General remarks on Mellin transform asymptotics}\label{se:mellin}

As mentioned after Lemma~\ref{le:dig slope}, we need the second order term
for the ATM digital call if we want to find the limiting slope in L\'evy
models with a Brownian component.
While this is easy for finite activity models (see the end of the preceding section),
it is more difficult in the case of infinite activity jumps.
We will find this second order term using Mellin transform
asymptotics.
For further details and references on this technique, see e.g.~\cite{FlGoDu95}.
The Mellin transform of a function~$H$, locally integrable on $(0,\infty)$, is defined by
\[
  (\mathcal{M}H)(s) = \int_0^\infty T^{s-1}H(T) dT.
\]
Under appropriate growth conditions on~$H$ at zero and infinity, this integral
defines an analytic function in an open vertical strip of the complex plane.
The function~$H$ can be recovered from its transform by Mellin inversion
(see formula~(7) in~\cite{FlGoDu95}):
\begin{equation}\label{eq:m inv}
  H(T) = \frac{1}{2\pi i}\int_{\kappa-i\infty}^{\kappa+i\infty} (\mathcal{M}H)(s) T^{-s} ds,
\end{equation}
where $\kappa$ is a real number in the strip of analyticity of~$\mathcal{M}H$.
For the validity of~\eqref{eq:m inv},
it suffices that~$H$ is continuous and that $y\mapsto (\mathcal{M}H)(\kappa+i y)$ is integrable.
Denote by~$s_0\in\mathbb{R}$ the real part of the left boundary of the strip of analyticity.
A typical situation in applications is that $\mathcal{M}H$ has a pole at~$s_0$,
and admits a meromorphic extension to a left half-plane, with further poles at
$s_0>s_1>s_2>\dots$ Suppose also that the meromorphic continuation
satisfies growth estimates at $\pm i\infty$ which allow to shift
the integration path in~\eqref{eq:m inv} to the left. We then collect the contribution
of each pole by the residue theorem, and arrive at an expansion
(see formula~(8) in~\cite{FlGoDu95})
\[
  H(T) = \mathrm{Res}_{s=s_0}(\mathcal{M}H)(s) T^{-s}
    + \mathrm{Res}_{s=s_1}(\mathcal{M}H)(s) T^{-s} + \dots
\]
Thus, the basic principle is that singularities~$s_i$ of the transform are mapped to
terms $T^{-s_i}$ in the asymptotic expansion of~$H$ at zero. Simple poles of $\mathcal{M}H$ 
yield powers of~$T$, whereas double poles produce an additional logarithmic factor
$\log T$, as seen from
the expansion $T^{-s}=T^{-s_i}(1-(\log T)(s-s_i)+O((s-s_i)^2))$.

\section{Main results: digital call prices and slope asymptotics}\label{se:levy mellin}

The mgf~$M(z,T)$ of~$X_T$ is analytic in a strip $z_- < \Re(z) < z_+$, given by
the critical moments
\begin{equation}\label{eq:s+}
  z_+ = \sup \{ z \in \mathbb{R} : \ev[e^{zX_T}] < \infty \}
\end{equation}
and
\begin{equation}\label{eq:s-}
  z_- = \inf \{ z \in \mathbb{R} : \ev[e^{zX_T}] < \infty \}.
\end{equation}
Since~$X$ is a L\'evy process, the critical moments do not depend on~$T$.
We will obtain asymptotic information on the transition probabilities (i.e.,
digital call prices) from the Fourier representation~\cite{Le04b}
\begin{align}
 \pp[S_T \geq 1]  &= \pp[X_T \geq 0] \notag \\
  &= \frac{1}{2i\pi}
   \int_{a-i\infty}^{a+i\infty}\frac{M(z,T)}{z} dz \notag \\
   &= \frac{1}{\pi}\Re \int_0^\infty \frac{M(a+iy,T)}{a+iy}dy, \label{eq:dig fourier}
\end{align}
where the real part of the vertical integration contour satisfies $a\in(0,1) \subseteq (z_-,z_+)$,
and convergence of the integral is assumed throughout.
We are going to analyze the asymptotic behavior of this integral, for $T\downarrow0$,
by computing its Mellin transform. Asymptotics of the probability (digital price)
$\pp[X_T \geq 0]$ are then evident from~\eqref{eq:dig fourier}.
The linearity of $\log M$ as a function of~$T$ enables us to evaluate
the Mellin transform in semi-explicit form. 
\begin{lemma}\label{le:mellin}
  Suppose that $S=e^X$ is a martingale, and that $\sigma>0$. Then, for any $a\in(0,1)$,
  the Mellin transform of the function
  \begin{equation}\label{eq:H}
    H(T) := \int_0^\infty \frac{e^{T\psi(a+iy)}}{a+iy}dy, \quad T>0,
  \end{equation}
  is given by
  \begin{equation}\label{eq:Ga F}
    (\mathcal{M}H)(s) = \Gamma(s) F(s),  \quad 0<\Re(s)<\tfrac12,
   \end{equation}
  where
  \begin{equation}\label{eq:F}
     F(s) = \int_0^\infty \frac{(-\psi(a+iy))^{-s}}{a+iy}dy , \quad 0<\Re(s)<\tfrac12.
  \end{equation}
   Moreover, $|(\mathcal{M}H)(s)|$ decays exponentially, if $\,\Re(s)\in(0,\tfrac12)$ is
   fixed and $|\Im(s)|\to\infty$.
\end{lemma}
See the appendix for the proof of Lemma~\ref{le:mellin}.
With the Mellin transform in hand, we now proceed to convert an expansion of
the mgf at $i\infty$ to an expansion of $\pp[X_T \geq 0]$ for $T\downarrow0$.
The following result covers, e.g., the NIG and Meixner models, and the generalized
tempered stable model, all with $\sigma>0$. See Section~\ref{se:ex} for details.
\begin{thm}\label{thm:dig2}
  Suppose that $S=e^X$ is a martingale, and that $\sigma>0$.
   Assume further that there are constants $a\in(0,1)$, $c\in\mathbb{C}$, $\nu\in[1,2)$ and
   $\varepsilon>0$ such that the Laplace exponent satisfies
  \begin{equation}\label{eq:psi2}
    \psi(z) = \frac12\sigma^2z^2 + cz^\nu + O(z^{\nu-\varepsilon}),\quad \Re(z)=a,\ \Im(z)\to\infty.
  \end{equation}
   Then the ATM digital call price satisfies 
  \begin{equation}\label{eq:dig2}
    \pp[X_T\geq 0] = \frac12 + C_{\tilde\nu}T^{\tilde\nu} + o(T^{\tilde\nu}),
    \quad T\downarrow0, 
  \end{equation}
  where  $C_{\tilde\nu}=\frac{\tilde\nu}{2\pi}
    \left(\tfrac12\sigma^2\right)^{\tilde\nu-1} \Im(e^{-i\pi\tilde\nu}c) \Gamma(-\tilde\nu)$
  with $\tilde{\nu}=(2-\nu)/2\in(0,\tfrac12]$.  
  For $\nu=1$, this simplifies to
   \begin{equation*}
    \pp[X_T\geq 0] = \frac12 + \frac{\Re(c)}{\sigma\sqrt{2\pi}} \sqrt{T} + o(\sqrt{T}),
    \quad T\downarrow0.
  \end{equation*}
\end{thm}
Together with Lemma~\ref{le:dig slope}, this theorem implies the following corollary,
which is our main result on the implied volatility slope as $T\downarrow0$.
\begin{cor}\label{cor:main}
 Under the assumptions of Theorem~\ref{thm:dig2}, the ATM implied volatility slope behaves as follows:
 \begin{itemize}
    \item [(i)] If $\nu=1$, then
       \[
         \lim_{T\downarrow0}\partial_K \sigma_{\mathrm{imp}}(K,T)|_{K=1} 
           = -\frac{\Re(c)}{\sigma}-\frac{\sigma}{2},
       \]
       with $c$ from~\eqref{eq:psi2}.
     \item [(ii)] If $1<\nu<2$ and $C_{\tilde{\nu}}\neq0$, then
      \[
         \partial_K \sigma_{\mathrm{imp}}(K,T)|_{K=1}  \sim -\sqrt{2\pi} C_{\tilde{\nu}}
             T^{\tilde{\nu}-1/2}, \quad T\downarrow0.
      \]
  \end{itemize}
\end{cor}
\begin{proof}[Proof of Theorem~\ref{thm:dig2}]
  From~\eqref{eq:dig fourier} and~\eqref{eq:H} we know that
  \begin{equation}\label{eq:P H}
    \pp[X_T\geq 0] = \frac{1}{\pi} \Re H(T).
  \end{equation}
  We now express $H(T)$ by the Mellin inversion formula~\eqref{eq:m inv}, with $\kappa\in(0,\tfrac12)$.
  This is justified by Lemma~\ref{le:mellin}, which yields the exponential decay of the transform
  $\mathcal{M}H$ along vertical rays. (Continuity of~$H$, which is also needed for the inverse transform,
  is clear.) Therefore, we have
  \begin{equation}\label{eq:H inv}
    H(T) = \frac{1}{2\pi i}\int_{1/4-i\infty}^{1/4+i\infty} \Gamma(s)F(s) T^{-s} ds, \quad T\geq 0.
  \end{equation}
  As outlined in Section~\ref{se:mellin}, we now show that $\Gamma(s)F(s)$ has a meromorphic
  continuation, then shift the integration path in~\eqref{eq:H inv} to the left, and collect residues.
  It is well known that~$\Gamma$ is meromorphic with poles at the non-positive integers, so it suffices
  to discuss the continuation of~$F$, defined in~\eqref{eq:F}. As in the proof of Lemma~\ref{le:mellin},
  we put $h(y):=-\psi(a+iy)$, $y\geq0$. To prove exponential decay of the desired meromorphic
  continuation, it is convenient to split the integral:
  \begin{align}
     F(s) &= \int_0^{y_0} \frac{h(y)^{-s}}{a+iy} dy 
     + \int_{y_0}^\infty \frac{h(y)^{-s}}{a+iy} dy \label{eq:Ft} \\
     &=: A_0(s) +\tilde{F}(s), \quad 0<\Re(s)<\tfrac12. \notag
  \end{align}
  The constant $y_0\geq0$ will be specified later. It is easy to see that~$A_0$ is analytic
  in the half-plane $\Re(s)<\tfrac12$, and so~$\tilde{F}$ captures all poles of~$F$ in that
  half-plane.
  By~\eqref{eq:psi2}, the function~$h$ has the
  expansion (with a possibly decreased~$\varepsilon$, to be precise)
  \begin{equation}\label{eq:h exp 1}
    h(y) = \tfrac12 \sigma^2 y^2 + \tilde{c}y^\nu + O(y^{\nu-\varepsilon}), \quad
     y\to\infty,
  \end{equation}
  where
  \[
    \tilde{c} :=
    \begin{cases}
      -c i^\nu & \nu > 1, \\
      -(c+\sigma^2 a)i & \nu=1.
    \end{cases}
  \]
  The reason why~$F$ (or $\tilde{F}$) is not analytic at $s=0$ is that the second
  integral in~\eqref{eq:Ft} fails to converge for $y$ large. 
  We thus subtract the following convergence-inducing integral from~$\tilde{F}$:
  \begin{align}
    \tilde{G}_1(s) &:= \int_{y_0}^\infty \frac{(\tfrac12 \sigma^2 y^2)^{-s}}{a+iy}dy \notag \\
    &= - \pi i (\tfrac12 a^2 \sigma^2)^{-s} \frac{e^{ i \pi s}}{\sin 2\pi s}
      - \int_{0}^{y_0} \frac{(\tfrac12 \sigma^2 y^2)^{-s}}{a+iy}dy \label{eq:G1} \\
    &=: G_1(s) + A_1(s). \notag
  \end{align}
  Note that $G_1$ is meromorphic, and that~$A_1$ is analytic for $\Re(s)<\tfrac12$.
  {}From the expansion
  \begin{equation}\label{eq:h exp}
    h(y)^{-s} = (\tfrac12 \sigma^2 y^2)^{-s} - \frac{2\tilde{c}s}{\sigma^2}\left(\frac{\sigma^2}{2}
    \right)^{-s} y^{\nu-2s-2} + O(y^{\nu-2\Re(s)-2-\varepsilon}), \quad y\to\infty,
  \end{equation}
  for $s$ fixed, we see that the function
  \begin{equation}\label{eq:Ft1}
    \tilde{F}_1(s)
      := \int_{y_0}^\infty  \frac{1}{a+iy} \left(h(y)^{-s} -  (\tfrac12 \sigma^2 y^2)^{-s}\right) dy
  \end{equation}
  is analytic for $-\tilde{\nu}<\Re(s)<\tfrac12$, and, clearly,
  for $0<\Re(s)<\tfrac12$ we have
  \begin{equation}\label{eq:FG}
    \tilde{F}(s) = \tilde{F}_1(s) + \tilde{G}_1(s).
  \end{equation}
  We have thus established the meromorphic continuation of~$\tilde{F}$
  to the strip $-\tilde{\nu}<\Re(s)<\tfrac12$. To continue~$\tilde{F}$ even further,
  we look at the second term in~\eqref{eq:h exp} and define
  \begin{align*}
    \tilde{G}_2(s) &:=  -\frac{2\tilde{c}s}{\sigma^2}\left(\frac{\sigma^2}{2}\right)^{-s}
    \int_{y_0}^\infty \frac{y^{\nu-2s-2}}{a+iy}dy \\
    &= -\frac{2\tilde{c}\pi}{\sigma^2}\left(\frac{\sigma^2}{2}\right)^{-s}s
    a^{\nu-2s-2}
     \frac{e^{(2s-\nu+3) \pi i/2}}{\sin \pi(\nu -2s)} 
      +\frac{2\tilde{c}s}{\sigma^2}\left(\frac{\sigma^2}{2}\right)^{-s}
    \int_0^{y_0} \frac{y^{\nu-2s-2}}{a+iy}dy \\
    &=: G_2(s) + A_2(s)
  \end{align*}
  and the compensated function
  \[
    \tilde{F}_2(s) := \int_{y_0}^\infty  \frac{1}{a+iy} \left(h(y)^{-s} -  (\tfrac12 \sigma^2 y^2)^{-s}
    +\frac{2\tilde{c}s}{\sigma^2}\left(\frac{\sigma^2}{2}\right)^{-s}y^{\nu-2s-2}\right) dy.
  \]
  By~\eqref{eq:h exp}, the function~$\tilde{F}_2$ is analytic for $\Re(s) \in
    (-\tilde{\nu}-\varepsilon/2,(\nu-1)/2) $. Moreover, by definition we have
  \[
    \tilde{F_1}(s) = \tilde{F}_2(s) + \tilde{G}_2(s), \quad -\tilde{\nu}
      < \Re(s) < \tfrac{\nu-1}{2},
  \]
  and so the meromorphic continuation of~$\tilde{F}$ to the region $-\tilde{\nu}-\varepsilon/2
  <\Re(s) < \tfrac12$ is established.
  
  In order to shift the integration path in~\eqref{eq:H inv} to the left, we have to ensure
  that the integral converges. This is the content
  of Lemma~\ref{le:decay} below, which also yields the existence
  of an appropriate $y_0\geq0$, to be used in the definition of~$\tilde{F}$ in~\eqref{eq:Ft}.
  By the residue theorem, we obtain
  \begin{multline}\label{eq:H res}
    H(T) = \mathrm{Res}_{s=0}(\mathcal{M}H)(s)T^{-s} 
    + \mathrm{Res}_{s=-\tilde{\nu}}(\mathcal{M}H)(s)T^{-s} \\
    + \frac{1}{2\pi i}\int_{\kappa-i\infty}^{\kappa+i\infty} (\mathcal{M}H)(s) T^{-s} ds, \quad T\geq0,
  \end{multline}
  where $\kappa=-\tilde{\nu}-\varepsilon/4$, and
  $\mathcal{M}H$ now of course denotes the meromorphic continuation of the Mellin transform.
  We then compute the residues. According to~\eqref{eq:Ft}  and~\eqref{eq:FG}, the continuation
  of $\mathcal{M}H$ in a neighborhood of $s=0$ is given by $\Gamma(s)(A_0(s)
  +\tilde{F}_1(s) + \tilde{G}_1(s))$. Therefore,
  \begin{align}
     \mathrm{Res}_{s=0}(\mathcal{M}H)(s)T^{-s} &= A_0(0)+\tilde{F}_1(0)+A_1(0)
       +\mathrm{Res}_{s=0}\Gamma(s)G_1(s)T^{-s} \notag \\
      &= \mathrm{Res}_{s=0}\Gamma(s)G_1(s)T^{-s} \label{eq:res}\\
    &= \tfrac12 \pi + i(\tfrac12\gamma - \log(a\sigma/\sqrt{2}) + \tfrac12 \log T), \notag
  \end{align}
  where~$\gamma$ is Euler's constant. Note that $A_0(0)=-A_1(0)$ and $\tilde{F_1}(0)=0$ by
  definition.
  The remaining residue~\eqref{eq:res} is straightforward to compute
  from~\eqref{eq:G1} (with a computer algebra system, e.g.)
  and has real part~$\tfrac12 \pi$. Notice that the logarithmic term $\log T$, resulting
  from the \emph{double} pole at zero (see the end of Section~\ref{se:mellin}),
  appears only in the imaginary part. Recalling~\eqref{eq:P H},
  we see that the first term
  on the right-hand side of~\eqref{eq:H res} thus yields the first term of~\eqref{eq:dig2}.
  
  Similarly, we compute for $\nu>1$
  \begin{align*}
    \mathrm{Res}_{s=-\tilde{\nu}}(\mathcal{M}H)(s)T^{-s}
      &= \mathrm{Res}_{s=-\tilde{\nu}} \Gamma(s)G_2(s)T^{-s} \\
      &=  \frac{\Gamma(-\tilde{\nu})}{2\pi}\left[
       \frac{2\tilde{c}s}{\sigma^2}\left(\frac{\sigma^2}{2}\right)^{-s}\pi a^{\nu-2s-2}
       e^{(2s-\nu+3)\pi i/2} T^{-s}
       \right]_{s=-\tilde{\nu}} .
  \end{align*}
  In the case $\nu=1$, the function $G_1$ also has a pole at $-\tilde{\nu}=-\tfrac12$,
  and we obtain
\begin{align*}
    \mathrm{Res}_{s=-\tilde{\nu}}(\mathcal{M}H)(s)T^{-s}
      &= \mathrm{Res}_{s=-1/2} \Gamma(s)(G_1(s)+G_2(s))T^{-s} \\
      &=     \sqrt{\frac{\pi}{2}}\left(\frac{i \tilde{c}}{\sigma}-a \sigma\right)\sqrt{T}.
  \end{align*}
  A straightforward computation shows that the stated formula for~$C_{\tilde{\nu}}$
  is correct in both cases.
  The integral on the right-hand side of~\eqref{eq:H res} is clearly
  $O(T^{-\kappa})=o(T^{\tilde{\nu}})$,
  and so the proof is complete.
\end{proof}
\begin{lemma}\label{le:decay}
  There is $y_0\geq0$ such that the meromorphic continuation of $\mathcal{M}H$ constructed in the proof of 
  Theorem~\ref{thm:dig2}, which depends on~$y_0$ via the definition of
  $\tilde{F}$ in~\eqref{eq:Ft}, decays exponentially as $|\Im(s)|\to\infty$.
\end{lemma}
Lemma~\ref{le:decay} is proved in the appendix.

\section{Examples}\label{se:ex}

\begin{figure}
\begin{center}
\includegraphics[height=1.8in]{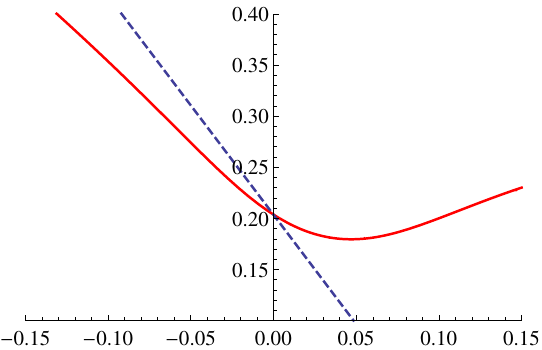}
\includegraphics[height=1.8in]{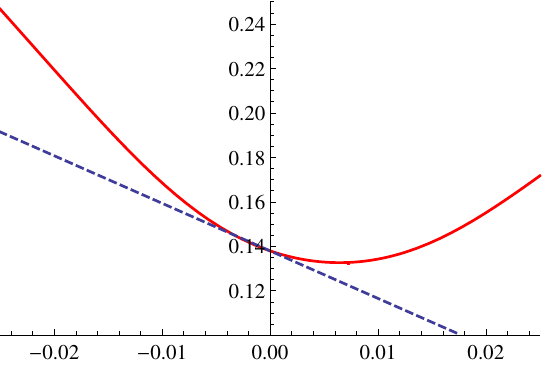}
\caption{\label{fig:nig}
The volatility smile, as a function of log-strike,
of the NIG model with parameters $\sigma=0.085$, $\hat\alpha=4.237$,
$\beta=-3.55$, $\delta=0.167$, and maturity $T=0.1$ (left panel) respectively
$T=0.01$ (right panel). The parameters were calibrated to S\&P 500 call prices
from Appendix~A of~\cite{Bu07}. The dashed line is the slope approximation~\eqref{eq:nig sl}.
We did the calibration and the plots with
Mathematica, using the Fourier representation of the call price.
}
\end{center}
\end{figure}

We now apply our main results (Theorem~\ref{thm:dig2} and Corollary~\ref{cor:main})
to several concrete models.
\begin{exm}\label{ex:nig}
The NIG (Normal Inverse Gaussian) model has Laplace exponent
\[
  \psi(z) = \tfrac12 \sigma^2z^2+\mu z+ \delta  (\sqrt{\hat{\alpha}^2-\beta^2}
    - \sqrt{\hat{\alpha}^2-(\beta+z)^2}),
\]
where $\delta>0$, $\hat{\alpha} > \max\{\beta+1,-\beta \}$. (The notation $\hat{\alpha}$
should avoid confusion with $\alpha$ from Theorem~\ref{thm:dig}.)
Since $S$ is a martingale, we must have
\[
  \mu = -\tfrac12 \sigma^2 + \delta(\sqrt{\hat{\alpha}^2-(\beta+1)^2} - \sqrt{\hat{\alpha}^2-\beta^2}).
\]
The relation between $\mu$ and $b$ from~\eqref{eq:psi} is
$\mu+\beta \delta/\sqrt{\hat\alpha^2-\beta^2}=b$, as seen from the derivative of the Laplace
exponent~$\psi$ at $z=0$.
The L\'evy density is
\[
\frac{\nu(dx)}{dx} = \frac{\delta \hat{\alpha}}{\pi |x|} e^{\beta x } K_1\bigl(\hat{\alpha} |x| \bigr),
\] 
where $K_1$ is the modified Bessel function of second order and index 1.

First assume $\sigma=0$.
Since $K_1(x) \sim 1/x$ for $x \downarrow 0$, condition $(\mathbf{H}$-$\alpha)$ is
satisfied with $\alpha=1$, with $c_+=c_-= \delta/\pi$. The integrability condition
in part~(iv) of Theorem~\ref{thm:dig} is easily checked, and we conclude
\[
  \lim_{T \downarrow 0} \pp[X_T \geq 0] = \frac 12+ \frac {1} {\pi}\arctan \Bigl(\frac {\mu} {\delta} \Bigr),
  \quad \sigma=0.
\]
Note that $b^*=\mu=b-\tfrac{\delta\hat\alpha}{\pi}\int_0^\infty
  K_1(\hat\alpha x)(e^{\beta x}-e^{-\beta x})dx$.
By Lemma~\ref{le:dig slope}, the implied volatility slope of the NIG model thus satisfies
\[
    \partial_K \sigma_{\mathrm{imp}}(K,T)|_{K=1} \sim
      -\sqrt{2/\pi} \arctan (\mu/\delta) \cdot T^{-1/2},
    \quad T\downarrow0, \quad \sigma=0,\ \mu\neq0.
\]
Now assume that $\sigma>0$. Since $\sqrt{\hat{\alpha}^2-(\beta+z)^2}=-iz + O(1)$
as $\Im(z)\to\infty$, the expansion~\eqref{eq:psi2} becomes
\[
  \psi(z) = \tfrac12\sigma^2z^2 + (\mu+ i)z + O(1), \quad \Re(z)=a,\ \Im(z)\to\infty.
\]
We can thus apply Theorem~\ref{thm:dig2} to conclude that
the ATM digital price satisfies
  \[
    \mathbb{P}[X_T \geq 0] = \frac12 + \frac{\mu}{\sigma \sqrt{2\pi}} \sqrt{T}
      + o(\sqrt{T}), \quad T\downarrow 0, \quad \sigma>0.
  \]
  By part (i) of Corollary~\ref{cor:main}, the limit of the implied volatility slope is given by
  \begin{align}
    \lim_{T\downarrow0}
    \partial_K \sigma_{\mathrm{imp}}(K,T)|_{K=1} &= -\frac{\mu}{\sigma}-\frac{\sigma}{2} \notag \\
    &= \frac{\delta}{\sigma}(\sqrt{\hat\alpha^2-\beta^2}
      -\sqrt{\hat\alpha^2-(\beta+1)^2} ), \quad \sigma>0. \label{eq:nig sl}
  \end{align}
  This limit is positive if and only if $\beta>-\tfrac12$.
\end{exm}
See Figure~\ref{fig:nig} for a numerical example. Let us stress again that
we identify the correct \emph{sign} of the slope, while we find that explicit asymptotics
do not approximate the \emph{value} of the slope very accurately. Still, in the right
panel of Figure~\ref{fig:nig} we have zoomed in at very short maturity to show
that our approximation gives the asymptotically correct tangent in this example.
\begin{exm}
The Laplace exponent of the Meixner model is
\[
  \psi(z) = \tfrac12\sigma^2z^2+ \mu z + 2\hat{d} \log \frac{\cos(\hat{b}/2)}{
  \cosh{\tfrac12(-\hat{a}iz-i\hat{b})}} ,
\]
where $\hat{d}>0$, $\hat{b}\in (-\pi,\pi)$,
and $0<\hat{a}<\pi - \hat{b}$.
(We follow the notation of Schoutens~\cite{Sc02}, except that we write~$\mu$ instead
of~$m$, and $\hat{a},\hat{b},\hat{d}$ instead of $a,b,d$.) The L\'evy density is
\[
\frac{\nu(dx)}{dx} = \hat{d} \frac{\exp(\hat{b}x/\hat{a})}{x \sinh(\pi x/\hat{a})}.
\] 
We can proceed analogously to Example~\ref{ex:nig}.
For $\sigma=0$ we again apply part~(iv) of Theorem~\ref{thm:dig}, with $\alpha=1$, where now
$c_+=c_-=\hat{d}\hat{a}/\pi$. Consequently,
\[
  \lim_{T\downarrow0}\pp[X_T\geq0] =  \frac12 + \frac{1}{\pi}
    \arctan\left(\frac{\mu}{\hat{a}\hat{d}}\right), \quad \sigma=0,
\]
and
\[
    \partial_K \sigma_{\mathrm{imp}}(K,T)|_{K=1} \sim
      -\sqrt{2/\pi} \arctan \left(\frac{\mu}{\hat{a}\hat{d}}\right) \cdot T^{-1/2},
    \quad T\downarrow0, \quad \sigma=0,\ \mu\neq0.
\]
Now assume $\sigma>0$. The expansion of the Laplace exponent is
\[
  \psi(z) = \tfrac12\sigma^2z^2 + (\mu+\hat{a}\hat{d} i)z + O(1), \quad \Re(z)=a,\ \Im(z)\to\infty.
\]
By Theorem~\ref{thm:dig2},
the ATM digital price in the Meixner model thus satisfies
  \[
    \mathbb{P}[X_T \geq 0] = \frac12 + \frac{\mu}{\sigma \sqrt{2\pi}} \sqrt{T}
      + o(\sqrt{T}), \quad T\downarrow 0.
  \]
  The limit of the implied volatility slope is given by
  \begin{align*}
    \lim_{T\downarrow0}
    \partial_K \sigma_{\mathrm{imp}}(K,T)|_{K=1} &= -\frac{\mu}{\sigma}-\frac{\sigma}{2} \\
    &= \frac{2\hat{d}}{\sigma} \log \left(\frac{\cos(\hat{b}/2)}{
       \cosh{\tfrac12(-(\hat{a}+\hat{b})i)}}\right), \quad \sigma>0.   
  \end{align*}
\end{exm}
\begin{exm}\label{ex:cgmy}
  The Laplace exponent of the CGMY model is
  \begin{equation}\label{eq:psi cgmy} 
    \psi(z) = \tfrac12\sigma^2 z^2 + \mu z + C\Gamma(-Y)((M-z)^Y - M^Y
    + (G+z)^Y - G^Y),
  \end{equation}
  where we assume $C>0$, $G>0$, $M>1$, $0<Y<2$, and $Y\neq1$.
  
  The case $\sigma=0$ and $Y\in(0,1)$ need not be discussed, as it is
  a special case of Proposition~8.5 in~\cite{AnLi13}. Our Proposition~\ref{prop:fin var}
  could also be applied, as the CGMY process has finite variation in this case.
  
  If  $\sigma=0$ and $Y\in(1,2)$,
  then the ATM digital call price converges to $\tfrac12$, and the slope explodes,
  of order $T^{1/2-1/Y}$. This is a special case of Corollary~3.3
  in~\cite{FiOl15}. Note that Proposition~8.5 in~\cite{AnLi13} is not applicable
  here,  because the constant $C_\mathfrak{M}$ from this proposition vanishes for the
  CGMY model, and so the leading term of the slope is not obtained.
  Theorem~\ref{thm:dig} (iv) from our Section~\ref{se:dig} is not useful, either;
  it gives the correct digital call limit price $\tfrac12$, but does not provide
  the second order term necessary to get slope asymptotics.
  
  We now proceed to the case $\sigma>0$, which is our main focus.
  The expansion of~$\psi$ at $i\infty$ is
  \begin{align*}
    \psi(z) = \tfrac12\sigma^2 z^2 + c_Yz^Y + \mu z + O(z^{Y-1}),\quad\Re(z)=a,\ \Im(z)\to\infty,
  \end{align*}
  with the complex constant $c_Y:=C\Gamma(-Y)(1+e^{-i\pi Y})$.
  First assume $0<Y<1$. Then we proceed analogously to the preceding examples,
  applying Theorem~\ref{thm:dig2} and Corollary~\ref{cor:main}.
  The ATM digital price thus satisfies
  \begin{equation}\label{eq:cgmy dig}
    \mathbb{P}[X_T \geq 0] = \frac12 + \frac{\mu}{\sigma \sqrt{2\pi}} \sqrt{T}
      + o(\sqrt{T}), \quad T\downarrow 0,
  \end{equation}
  and the limit of the implied volatility slope is given by
  \begin{align}
    \lim_{T\downarrow0}
    \partial_K \sigma_{\mathrm{imp}}(K,T)|_{K=1} &= -\frac{\mu}{\sigma}-\frac{\sigma}{2} \notag \\
    &= \frac{1}{\sigma}C\Gamma(-Y)((M-1)^Y - M^Y + (G+1)^Y - G^Y). \label{eq:cgmy sl}
  \end{align}
  Now assume $1<Y<2$.  
  In principle, Theorem~\ref{thm:dig2} is applicable, with $\nu=Y$;
  however, the constant $C_{\tilde{\nu}}$ in~\eqref{eq:dig2} is zero,
  and so we do not get the second term of the expansion immediately.
  What happens is that the Mellin
  transform of~$H$ (see the proof of Theorem~\ref{thm:dig2}) may have further poles
  in $-\tfrac12<\Re(s)<0$, but none of them gives a contribution, since the corresponding
  residues have zero real part. Therefore,
  \eqref{eq:cgmy dig} and~\eqref{eq:cgmy sl} are true also for $1<Y<2$.  See
  A.~Pinter's forthcoming PhD thesis for details.
  Note that~\eqref{eq:cgmy dig} and~\eqref{eq:cgmy sl} also follow from concurrent
  work by Figueroa-L\'opez and \'Olafsson~\cite{FiOl15}.
  For $0<Y<1$, they also follow from Proposition~8.5 in~\cite{AnLi13}, but not
  for $1<Y<2$, because then the constant~$C_{\mathfrak{M}}$ from that proposition vanishes
  when specializing it to the CGMY model.
\end{exm}
In the following example, we discuss
the generalized tempered stable model. The tempered stable model, which is investigated
in~\cite{AnLi13}, is obtained by setting $\alpha_-=\alpha_+$.
\begin{exm}
The generalized tempered stable process~\cite{CoTa04} is a generalization of the CGMY model, with L\'evy density
\[
\frac{\nu(dx)}{dx}=  \frac{C_-}{|x|^{1+\alpha_-}}e^{-\lambda_-|x|}1_{(-\infty,0)}(x)
  + \frac{C_+}{|x|^{1+\alpha_+}}e^{-\lambda_+|x|}1_{(0,\infty)}(x),
\]
where $\alpha_\pm<2$ and $C_\pm,\lambda_\pm>0$.
For $\alpha_\pm\not\in\{0,1\}$ the Laplace exponent of the generalized tempered stable process is
\begin{multline*}
  \psi(z) = \tfrac12\sigma^2 z^2 + \mu z + \Gamma(-\alpha_+)C_+\Big((\lambda_+ - z)^{\alpha_+} - \lambda_+^{\alpha_+}\Big) \\
  + \Gamma(-\alpha_-)C_-\Big((\lambda_- + z)^{\alpha_-} - \lambda_-^{\alpha_-}\Big).
\end{multline*}
For $\sigma>0$, $\alpha_+\in(1,2)$, and $\alpha_-<\alpha_+$ we have the following expansion:
\[
  \psi(z) = \tfrac12\sigma^2 z^2 + \Gamma(-\alpha_+)C_+e^{-i\pi\alpha_+}z^{\alpha_+}
  + O(z^{\max\{1,\alpha_-\}}), \quad \Re(z)=a,\ \Im(z)\to\infty.
\]
We now apply Theorem~\ref{thm:dig2} with $\nu=\alpha_+$, and find that
the second order expansion of the ATM digital call is
\[
  \pp[X_T\geq 0] = \frac12 + C_{\tilde\nu}T^{\tilde\nu} + o(T^{\tilde\nu}),
  \quad T\downarrow0,  
\]
with $\tilde\nu=1-\alpha_+/2\in(0,\tfrac12)$ and the real constant 
\[
  C_{\tilde\nu}=\frac{\tilde\nu}{2\pi} \left(\tfrac12\sigma^2\right)^{\tilde\nu-1} \Gamma(-\alpha_+)C_+\underbrace{\Im(e^{-i\pi\tilde\nu}e^{-i\pi\alpha_+})}_{=\sin(-\pi(1+\alpha_+/2))} \Gamma(-\tilde\nu).
\]
By Corollary~\ref{cor:main} (ii), the ATM implied volatility slope explodes, but
slower than $T^{-1/2}$:
\[
  \partial_K \sigma_{\mathrm{imp}}(K,T)|_{K=1} \sim
    - \sqrt{2\pi}C_{\tilde{\nu}}T^{\tilde{\nu}-1/2}, \quad T\downarrow0.
\]
Note that these results also follow from the concurrent paper~\cite{FiOl15},
which treats tempered stable-like models.

If $\sigma>0$ and $\alpha_+<1$, then part~(i) of Corollary~\ref{cor:main} is applicable, and
formulas analogous to~\eqref{eq:cgmy dig} and~\eqref{eq:cgmy sl} hold.
\end{exm}

\section{Robustness of Lee's Moment Formula}\label{se:lee}

As we have already mentioned, our first order slope approximations give limited
accuracy for the size of the slope, but usually succeed at identifying
its sign, i.e.,
whether the smile increases or decreases at the money. It is a natural
question whether this sign gives information on the smile as a whole:
If the slope is positive, does it follow that the right wing is steeper
than the left one, and vice versa? To deal with this issue, recall Lee's
moment formula~\cite{Le04a}.
Under the assumption that the critical moments~$z_+$ and~$z_-$,
defined in~\eqref{eq:s+} and~\eqref{eq:s-}, are finite, Lee's
formula states that
\begin{equation}\label{eq:lee1}
  \limsup_{k\to\infty} \frac{\sigma_{\mathrm{imp}}(K,T)}{\sqrt{k}}
    = \sqrt{\frac{\Psi(z_+ - 1)}{T}}
\end{equation}
and
\begin{equation}\label{eq:lee2}
  \limsup_{k\to-\infty} \frac{\sigma_{\mathrm{imp}}(K,T)}{\sqrt{-k}}
    = \sqrt{\frac{\Psi(-z_-)}{T}},
\end{equation}
where $T>0$ is fixed, $k=\log K$, and $\Psi(x):=2-4(\sqrt{x^2+x}-x)$.
According to Lee's formula, the slopes of the wings
depend on the size of the critical moments. In L\'evy models,
the critical moments do not depend on~$T$.
The compatibility property we seek now becomes:
  \begin{equation}\label{eq:lee}
    \lim_{k\to\infty} \frac{\sigma_{\mathrm{imp}}(K,T)}{\sqrt{k}} >
    \lim_{k\to-\infty} \frac{\sigma_{\mathrm{imp}}(K,T)}{\sqrt{-k}}
    \quad \text{for all}\ T>0
  \end{equation}
  if and only if
  \begin{equation}\label{eq:slope lim}
    \partial_K \sigma_{\mathrm{imp}}(K,T) |_{K=1} >0 \quad \text{
    for all sufficiently small}\ T.
  \end{equation}
  That is, the right wing of the smile is steeper than the left
  wing
  deep \emph{out-of-the-money} if and only if the small-maturity
  \emph{at-the-money} slope is positive.  
We now show that this is true for several infinite activity L\'evy models. By our methods,
this can certainly be extended to other infinite activity models.
It does not hold, though, for the Merton and Kou jump diffusion models.
The parameter ranges in the following theorem are the same
as in the examples in Section~\ref{se:ex}.
\begin{thm}
  Conditions~\eqref{eq:lee} and~\eqref{eq:slope lim} are equivalent for the following models.
  For the latter three, we assume that $\sigma>0$ or $\mu\neq0$.
  \begin{itemize}
     \item Variance gamma with $\sigma=0$, $b_0\neq0$
     \item NIG
     \item Meixner
    \item CGMY
  \end{itemize}
  \end{thm}
  Put differently,
these models are \emph{not} capable (at short maturity)
of producing a smile that has, say, its minimum to the left of
$\log K=k=0$, and thus a positive ATM slope,
but whose left wing is steeper than the right one.
\begin{proof}
  The critical moments are clearly finite for all of these models. Moreover,
  it is well known that the $\limsup$ in~\eqref{eq:lee1} and~\eqref{eq:lee2}
  can typically be replaced by a genuine limit, for instance using
  the criteria given by Benaim and Friz~\cite{BeFr08}. Their conditions
  on the mgf are easily verified for all our models; in fact
  Benaim and Friz~\cite{BeFr08} explicitly treat the variance gamma model
  with $b_0=0$ and the NIG model.  
  We thus have to show
  that~\eqref{eq:slope lim} is equivalent to $\Psi(z_+ - 1) > \Psi(-z_-)$.
  Since $\Psi$ is strictly decreasing on $(0,\infty)$, the latter condition
  is equivalent to $z_+-1 < -z_-$. It remains to check the equivalence
  \begin{equation}\label{eq:new eq}
    z_+-1 < -z_- \quad \Longleftrightarrow \quad \eqref{eq:slope lim}.
  \end{equation}
  The mgf of the variance gamma model is (see~\cite{MaCaCh98})
  \[
    M(z,T) = e^{Tb_0z}(1-\theta \nu z - \tfrac12 \hat\sigma^2 \nu z^2)^{-T/\nu},
  \]
  where $\hat\sigma,\nu>0$ and $\theta\in \mathbb{R}$. Its paths have finite variation,
  and so Proposition~\ref{prop:fin var} shows that~\eqref{eq:slope lim} is equivalent
  to $b_0<0$.
  The critical moments are
  \[
    z_\pm = -\frac{\nu\theta \pm \sqrt{2\nu \hat\sigma^2 + \nu^2\theta^2}}
    {\nu\hat\sigma^2},
  \]
  and we have $-z_-+1-z_+ = 1+2\theta/\hat\sigma^2$. This is positive
  if and only if
  \[
    b_0 = \nu^{-1}\log(1-\theta\nu - \tfrac12 \hat\sigma^2 \nu) < 0,
  \]
  which yields~\eqref{eq:new eq}.
  
  As for the other three models, first suppose that $\sigma>0$. The examples in Section~\ref{se:ex} show
  that~\eqref{eq:slope lim} is equivalent to $\mu<-\tfrac12\sigma^2$. 
  The critical moments of the NIG model are $z_+=\hat\alpha-\beta$
  and $z_- = -\hat\alpha-\beta$. Therefore, $z_+-1 < -z_-$ if and only if
  $\beta>-\tfrac12$, and this is indeed equivalent to
  \[
    \mu+\tfrac12\sigma^2 = \delta(\sqrt{\hat\alpha^2-(\beta+1)^2}-\sqrt{\hat\alpha^2-\beta^2}) < 0.
  \]
 
  For the Meixner model, we have $z_\pm
  =(\pm \pi - \hat{b})/\hat{a}$, which yields $-z_-+1-z_+ = 1+2\hat{b}/\hat{a}$.
  On the other hand,
  \[
    \mu+\tfrac12\sigma^2  = -2 \hat{d} \log \frac{\cos(\hat{b}/2)}{\cos((\hat{a}+\hat{b})/2)},
  \]
  which is negative if and only if $\cos(\hat{b}/2) > \cos((\hat{a}+\hat{b})/2)$,
  and this is equivalent to $\hat{a}+2\hat{b}>0$.
  
  Finally, in case of the CGMY model, we have
  \[
    \mu+\tfrac12\sigma^2 = -C \Gamma(-Y)\big((M-1)^Y - M^Y + (G+1)^Y -G^Y\big).
  \]
  Since, for $Y\in(0,1)$, $\Gamma(-Y)<0$ and the function $x\mapsto x^Y-(x+1)^Y$
  is strictly increasing on $(0,\infty)$, we see that $\mu+\tfrac12\sigma^2<0$ if and only
  if $M-1 < G$. This is the desired condition, since the explicit
  expression~\eqref{eq:psi cgmy} shows that $z_+=M$ and $z_-=-G$.
  The case  $Y\in(1,2)$ is analogous.
  
  It remains to treat the case $\sigma=0$. First, note that the critical moments
  do not depend on~$\sigma$. Furthermore,
  from the examples in Section~\ref{se:ex}, we see that~\eqref{eq:slope lim} holds
  if and only if $\mu < 0$.
  Now observe that adding a Brownian motion
  $\sigma W_t$ to a L\'evy model adds $-\tfrac12\sigma^2$ to the drift, if the martingale
  property is to be preserved. Therefore, the assertion
  follows from what we have already proved about $\sigma>0$.
\end{proof}

\section{Conclusion}

Our main result (Corollary~\ref{cor:main})
translates asymptotics of the log-underlying's mgf to first-order asymptotics for the ATM
implied volatility slope.
Checking the requirements
of Corollary~\ref{cor:main} only requires Taylor expansion of the mgf, which has an
explicit expression in all models of practical interest.
Higher order expansions can be obtained by the same
proof technique, if desired. They will follow in a relatively straightforward
way from higher order expansions of the mgf, by collecting further
residues of the Mellin transform. In future work, we hope
to connect our assumptions on the mgf with properties of the L\'evy triplet,
which should give additional insight on how the slope depends on model
characteristics.

\appendix

\section{Proofs of Lemmas~\ref{le:mellin} and~\ref{le:decay}}

\begin{proof}[Proof of Lemma~\ref{le:mellin}]
  Since $S=e^X$ is a martingale, we have $\psi'(0)=\ev[X_1]<0$. Then $\psi(0)=0$ implies that
  $\psi(a)<0$ for all sufficiently small $a>0$. In fact, it easily
  follows from $\psi(1)=0$ and the concavity of~$\psi$ that all $a\in(0,1)$ satisfy $\psi(a)<0$.
  Let us fix such an~$a$. From
  \[
    \Re(-\psi(a+iy))=-\psi(a) + \frac12\sigma^2y^2 
      + \int_{\mathbb{R}}e^{ax}\underbrace{(1-\cos(yx))}_{\geq 0}\nu(dx)
  \]
  we obtain that the function $h(y):=-\psi(a+iy)$, $y\geq0$, satisfies
  \begin{equation}\label{eq:h est}
    \Re h(y) > \tfrac12 \sigma^2 y^2 \geq0, \quad y\geq0. 
  \end{equation}
   For $0<\Re(s)<\tfrac12$ define the function
  \[
    g(T) = T^{\Re(s)-1}\int_{0}^\infty \frac{e^{-T\Re(h(y))}}{|a+iy|}dy, \quad T>0.
  \]
  Using Fubini's theorem and substituting $T\Re(h(y))=u$, we then calculate for $\Re(s)>0$
  \begin{align*}
     \int_0^\infty g(T) dT &=
       \int_{0}^\infty \frac{1}{|a+iy|} \int_0^\infty e^{-T\Re(h(y))} T^{\Re(s)-1} dTdy  \\
      &= \int_{0}^\infty \frac{\Re(h(y))^{-\Re(s)}}{|a+iy|} \left(\int_0^\infty e^{-u}u^{\Re(s)-1}du\right) dy\\
      &=  \Gamma(\Re(s)) \int_{0}^\infty \frac{\Re(h(y))^{-\Re(s)}}{|a+iy|}dy.
  \end{align*}
  From~\eqref{eq:h est}, we get
  \begin{align*}
    \int_{0}^\infty \frac{\Re(h(y))^{-\Re(s)}}{|a+iy|}dy
    \leq (\tfrac12\sigma^2)^{-\Re(s)} \int_{0}^\infty\frac{y^{-2\Re(s)}}{|a+iy|}dy.
  \end{align*}
  The restriction $\Re(s)<\frac12$ ensures that
  the last integral is finite and thus the integrability of $g$.
  Using the dominated convergence theorem and Fubini's theorem,
  the Mellin transform of $H$ can now be calculated as
  \[
     \int_0^\infty H(T) T^{s-1} dT
       = \int_{0}^\infty \frac{1}{a+iy} \int_0^\infty e^{-Th(y)} T^{s-1} dTdy.
  \]
  The substitution $Th(y)=u$ gives us the result. Note that $h(y)$ is in general non-real;
  it is easy to see, though, that Euler's integral
  \[
    \Gamma(s) = \int_0^\infty u^{s-1}e^{-u} du, \quad \Re(s)>0,
  \]
  still represents
  the gamma function if the integration is performed along any complex ray emanating from zero, as
  long as the ray stays in the right half-plane. The latter holds, since $\Re(h(y))>0$.
  
  It remains to prove the exponential decay of the Mellin transform $\mathcal{M}H(s)=\Gamma(s)F(s)$
  for large $|\Im(s)|$. First, note that
  \begin{align*}
    \Im \psi(a+iy) &= by+\sigma^2ay + \int_{\mathbb{R}}(e^{ax}\sin xy
      +xy)\nu(dx) \\
    &= O(y), \quad y\to\infty,
  \end{align*}
  which together with~\eqref{eq:h est} yields the existence of an $\varepsilon>0$
  such that $|\arg h(y)|\leq \tfrac12\pi - \varepsilon$ for all $y\geq0$.
  We then estimate, with $\Re(s)\in(0,\tfrac12)$ fixed,
  \begin{align*}
     |F(s)| & \leq \int_{0}^\infty \frac{e^{-\Re(s \log h(y))}}{|a+iy|}dy \\
     &= \int_{0}^\infty \frac{e^{-\Re(s) \log| h(y)| + \Im(s) \arg h(y)}}{|a+iy|}dy \\
     &\leq e^{(\pi/2-\varepsilon) |\Im(s)|}
       \int_{0}^\infty \frac{(\tfrac12 \sigma^2 y^2)^{-\Re(s)}}{|a+iy|} dy.
  \end{align*}
  The integral converges, and thus this estimate is good enough, since Stirling's formula yields
  $|\Gamma(s)| = \exp\big(-\tfrac12 \pi |\Im(s)|(1+o(1))\big)$.
\end{proof}
\begin{proof}[Proof of Lemma~\ref{le:decay}]
  Recall that, in the proof of Theorem~\ref{thm:dig2}, we defined
  the following meromorphic continuation of~$F(s)$, to the strip
  $-\tilde{\nu}-\tfrac12 \varepsilon<\Re(s)<\tfrac12$:
  \begin{align*}
    &A_0(s) + \tilde{G}_1(s) + \tilde{F}_1(s),  \quad -\tilde{\nu}<\Re(s)<\tfrac12, \\
    &A_0(s) + \tilde{G}_1(s) + \tilde{G}_2(s) + \tilde{F}_2(s), 
     \quad -\tilde{\nu}-\tfrac12 \varepsilon<\Re(s)<\tfrac12(\nu-1) .
  \end{align*}
  As noted at the end of the proof of Lemma~\ref{le:mellin}, Stirling's formula
  implies  $|\Gamma(s)| = \exp\big(-\tfrac12 \pi |\Im(s)|(1+o(1))\big)$.
  By~\eqref{eq:Ga F}, it thus suffices to argue that the continuation of~$F(s)$
  is $O(\exp((\tfrac12 \pi-\varepsilon)|\Im(s)|))$ for some $\varepsilon>0$.
  The functions~$\tilde{G}_1$ and~$\tilde{G}_2$ are clearly~$O(1)$.
  As for~$A_0$, defined in~\eqref{eq:Ft}, we have
  \begin{align*}
    |A_0(s)| &\leq \int_0^{y_0} \frac{e^{-\Re(s \log h(y))}}{|a+iy|}dy \\
    &= \int_0^{y_0}\frac{|h(y)|^{-\Re(s)} e^{\Im(s) \arg h(y)}}{|a+iy|}  dy.
  \end{align*}
  Now note that
  \[
    |h(y)|^{-\Re(s)} \leq
    \begin{cases}
      (\tfrac12 \sigma^2 y^2)^{-\Re(s)} & 0<\Re(s)<\tfrac12, \\
      \left( \max_{0\leq y\leq y_0}|h(y)| \right)^{-\Re(s)}& \Re(s)\leq0,
    \end{cases}
  \]
  and that
  \[
    \exp(\Im(s) \arg h(y)) \leq \exp((\tfrac{\pi}{2}-\varepsilon)|\Im(s)|)
  \]
  for some $\varepsilon>0$, as argued in the proof of Lemma~\ref{le:mellin}.
  
  It remains to establish a bound for~$\tilde{F}_1$, defined in~\eqref{eq:Ft1}.
  (The bound for~$\tilde{F}_2$ is completely analogous, and we omit the details.)
  In what follows, we assume that $-\tilde{\nu}<\Re(s)<\tfrac12$.
  By~\eqref{eq:h exp 1}, we have (where the $O$ is uniform w.r.t.~$s$, and $y_0\geq0$ is
  still arbitrary):
  \begin{align}
    \tilde{F}_1(s) &= 
    \int_{y_0}^\infty \frac{1}{a+iy}\left((\tfrac12 \sigma^2 y^2)^{-s}
    (1+O(y^{\nu-2}))^{-s} -  (\tfrac12 \sigma^2 y^2)^{-s} \right)dy \notag \\
    &= \int_{y_0}^\infty \frac{1}{a+iy}
    (\tfrac12 \sigma^2 y^2)^{-s}\left((1+O(y^{\nu-2}))^{-s}  - 1
    \right)dy. \label{eq:F1 est}
  \end{align}
  We now choose~$y_0$ such that, for some constant $C_0>0$,
  \begin{align*}
    \big|\log|1+O(y^{\nu-2})|\big|&\leq \tfrac14 \pi, \\
    \big|\operatorname{arg}(1+O(y^{\nu-2}))\big|&\leq \tfrac14 \pi, \\
    \big|\log(1+O(y^{\nu-2}))\big|&\leq C_0 y^{\nu-2},
  \end{align*}
  hold for all $y\geq y_0$.
  (By a slight abuse of notation, here $O(y^{\nu-2})$ of course denotes the function
  hiding behind the $O(y^{\nu-2})$ in~\eqref{eq:F1 est}.)
  For all $w\in\mathbb{C}$ we have the estimate 
  \[
    |e^w-1| \leq |w|e^{|\Re(w)|}.
  \]
  Using this in~\eqref{eq:F1 est}, we find
  \begin{align*}
    \left|(1+O(y^{\nu-2}))^{-s}  - 1\right| &=
      \left|\exp(-s\log(1+O(y^{\nu-2})))-1\right| \\
    &\leq   |s\log(1+O(y^{\nu-2}))|\cdot\exp(|\Re(s\log(1+O(y^{\nu-2}))|)\\
    &\leq C_1|s|y^{\nu-2}\exp(\tfrac14 \pi|\Im(s)|),
  \end{align*}
  where $C_1=C_0\exp(\tfrac14 \pi \sup_s|\Re(s)|)$, and thus
  \begin{align*}
    |\tilde{F}_1(s)| &\leq  C_2|s|e^{\tfrac14 \pi|\Im(s)|} \int_{y_0}^\infty y^{-2\Re(s)+\nu-3}\,dy \\
    &= \exp\big(\tfrac14 \pi|\Im(s)|(1+o(1))\big).
  \end{align*}
\end{proof}

\bibliographystyle{siam}
\bibliography{../gerhold}

\def\polhk#1{\setbox0=\hbox{#1}{\ooalign{\hidewidth
  \lower1.5ex\hbox{`}\hidewidth\crcr\unhbox0}}}
\begin{thebibliography}{10}

\bibitem{Ai02}
{\sc Y.~A{\"{\i}}t-Sahalia}, {\em Telling from discrete data whether the
  underlying continuous-time model is a diffusion}, Journal of Finance, 57
  (2002), pp.~2075--2113.

\bibitem{AiJa10}
{\sc Y.~A{\"{\i}}t-Sahalia and J.~Jacod}, {\em Is {B}rownian motion necessary
  to model high-frequency data?}, Ann. Statist., 38 (2010), pp.~3093--3128.

\bibitem{AlLeVi07}
{\sc E.~Al{\`o}s, J.~A. Le{\'o}n, and J.~Vives}, {\em On the short-time
  behavior of the implied volatility for jump-diffusion models with stochastic
  volatility}, Finance Stoch., 11 (2007), pp.~571--589.

\bibitem{AnLi13}
{\sc L.~Andersen and A.~Lipton}, {\em Asymptotics for exponential {L}\'evy
  processes and their volatility smile: survey and new results}, Int. J. Theor.
  Appl. Finance, 16 (2013).
\newblock Paper no.\ 1350001, 98 pages.

\bibitem{BaFrGa16}
{\sc C.~Bayer, P.~Friz, and J.~Gatheral}, {\em Pricing under rough volatility},
  Quantitative Finance, 16 (2016), pp.~887--904.

\bibitem{BeFr08}
{\sc S.~Benaim and P.~Friz}, {\em Smile asymptotics {II}: {M}odels with known
  moment generating functions}, J. Appl. Probab., 45 (2008), pp.~16--32.

\bibitem{BoLe02}
{\sc S.~I. Boyarchenko and S.~Z. Levendorski{\u\i}}, {\em Non-{G}aussian
  {M}erton-{B}lack-{S}choles theory}, vol.~9 of Advanced Series on Statistical
  Science \& Applied Probability, World Scientific Publishing Co., Inc., River
  Edge, NJ, 2002.

\bibitem{Bu07}
{\sc Y.~Bu}, {\em Option pricing using {L}{\'e}vy processes}, master's thesis,
  {C}halmers {U}niversity of {T}echnology, {G}{\"o}teborg, 2007.

\bibitem{CaWu03}
{\sc P.~Carr and L.~Wu}, {\em What type of process underlies options? {A}
  simple robust test}, The Journal of Finance, 58 (2003), pp.~2581--2610.

\bibitem{CoTa04}
{\sc R.~Cont and P.~Tankov}, {\em Financial modelling with jump processes},
  Chapman \& Hall/CRC Financial Mathematics Series, Chapman \& Hall/CRC, Boca
  Raton, FL, 2004.

\bibitem{DaIb71}
{\sc Y.~A. Davydov and I.~A. Ibragimov}, {\em On asymptotic behavior of some
  functionals of processes with independent increments}, Theory Probab. Appl.,
  16 (1971), pp.~162--167.

\bibitem{DeVaCiBo12}
{\sc L.~De~Leo, V.~Vargas, S.~Ciliberti, and J.-P. Bouchaud}, {\em We've walked
  a million miles for one of these smiles}.
\newblock Preprint, available at \url{http://arxiv.org/abs/1203.5703}, 2012.

\bibitem{DoMa02}
{\sc R.~A. Doney and R.~A. Maller}, {\em Stability and attraction to normality
  for {L}\'evy processes at zero and at infinity}, J. Theoret. Probab., 15
  (2002), pp.~751--792.

\bibitem{Du10}
{\sc V.~Durrleman}, {\em From implied to spot volatilities}, Finance Stoch., 14
  (2010), pp.~157--177.

\bibitem{Fa14}
{\sc J.~Fajardo}, {\em Barrier style contracts under {L}{\'e}vy processes: an
  alternative approach}.
\newblock Preprint, available on SSRN, 2014.

\bibitem{FaMo06}
{\sc J.~Fajardo and E.~Mordecki}, {\em Symmetry and duality in {L}\'evy
  markets}, Quant. Finance, 6 (2006), pp.~219--227.

\bibitem{FiFo12}
{\sc J.~E. Figueroa-L{\'o}pez and M.~Forde}, {\em The small-maturity smile for
  exponential {L}\'evy models}, SIAM J. Financial Math., 3 (2012), pp.~33--65.

\bibitem{FiGoHo12}
{\sc J.~E. Figueroa-L{\'o}pez, R.~Gong, and C.~Houdr{\'e}}, {\em Small-time
  expansions of the distributions, densities, and option prices of stochastic
  volatility models with {L}\'evy jumps}, Stochastic Process. Appl., 122
  (2012), pp.~1808--1839.

\bibitem{FiGoHo14}
{\sc J.~E. Figueroa-L{\'o}pez, R.~Gong, and C.~Houdr{\'e}}, {\em High-order
  short-time expansions for {ATM} option prices of exponential {L}{\'e}vy
  models}.
\newblock To appear in Mathematical Finance, 2014.

\bibitem{FiHo09}
{\sc J.~E. Figueroa-L{\'o}pez and C.~Houdr{\'e}}, {\em Small-time expansions
  for the transition distributions of {L}\'evy processes}, Stochastic Process.
  Appl., 119 (2009), pp.~3862--3889.

\bibitem{FiOl15}
{\sc J.~E. Figueroa-L{\'o}pez and S.~{\'O}lafsson}, {\em Short-time asymptotics
  for the implied volatility skew under a stochastic volatility model with
  {L}{\'e}vy jumps}.
\newblock Preprint, available at \url{http://arxiv.org/abs/1502.02595}, 2015.

\bibitem{FlGoDu95}
{\sc P.~Flajolet, X.~Gourdon, and P.~Dumas}, {\em Mellin transforms and
  asymptotics: harmonic sums}, Theoret. Comput. Sci., 144 (1995), pp.~3--58.
\newblock Special volume on mathematical analysis of algorithms.

\bibitem{FoJaFi11}
{\sc M.~Forde, A.~Jacquier, and J.~E. Figueroa-L{\'o}pez}, {\em The large-time
  smile and skew for exponential {L}{\'e}vy models}.
\newblock Preprint, 2011.

\bibitem{FrGeGuSt11}
{\sc P.~Friz, S.~Gerhold, A.~Gulisashvili, and S.~Sturm}, {\em On refined
  volatility smile expansion in the {H}eston model}, Quantitative Finance, 11
  (2011), pp.~1151--1164.

\bibitem{Ga06}
{\sc J.~Gatheral}, {\em {The Volatility Surface, A Practitioner's Guide}},
  Wiley, 2006.

\bibitem{Ge13}
{\sc S.~Gerhold}, {\em Can there be an explicit formula for implied
  volatility?}, Appl. Math. E-Notes, 13 (2013), pp.~17--24.

\bibitem{GeKlPoSh15}
{\sc S.~Gerhold, M.~Kleinert, P.~Porkert, and M.~Shkolnikov}, {\em Small time
  central limit theorems for semimartingales with applications}, Stochastics,
  87 (2015), pp.~723--746.

\bibitem{Le04a}
{\sc R.~W. Lee}, {\em The moment formula for implied volatility at extreme
  strikes}, Math. Finance, 14 (2004), pp.~469--480.

\bibitem{Le04b}
\leavevmode\vrule height 2pt depth -1.6pt width 23pt, {\em Option pricing by
  transform methods: Extensions, unification, and error control}, Journal of
  Computational Finance, 7 (2004), pp.~51--86.

\bibitem{Le05}
\leavevmode\vrule height 2pt depth -1.6pt width 23pt, {\em Implied volatility:
  statics, dynamics, and probabilistic interpretation}, in Recent advances in
  applied probability, Springer, New York, 2005, pp.~241--268.

\bibitem{MaCaCh98}
{\sc D.~Madan, P.~Carr, and E.~Chang}, {\em The variance gamma process and
  option pricing}, European Finance Review, 2 (1998), pp.~79--105.

\bibitem{MiTa12}
{\sc A.~Mijatovi{\'c} and P.~Tankov}, {\em A new look at short-term implied
  volatility in asset price models with jumps}.
\newblock Preprint, available at \url{http://arxiv.org/abs/1207.0843}, 2012.

\bibitem{Ro08}
{\sc M.~Roper}, {\em Implied volatility explosions: {E}uropean calls and
  implied volatilities close to expiry in exponential {L}{\'e}vy models}.
\newblock Preprint, available at \url{http://arxiv.org/abs/0809.3305}, 2008.

\bibitem{RoRu09}
{\sc M.~Roper and M.~Rutkowski}, {\em On the relationship between the call
  price surface and the implied volatility surface close to expiry}, Int. J.
  Theor. Appl. Finance, 12 (2009), pp.~427--441.

\bibitem{RoTa11}
{\sc M.~Rosenbaum and P.~Tankov}, {\em Asymptotic results for time-changed
  {L}\'evy processes sampled at hitting times}, Stochastic Process. Appl., 121
  (2011), pp.~1607--1632.

\bibitem{Sa99}
{\sc K.-i. Sato}, {\em L\'evy processes and infinitely divisible
  distributions}, vol.~68 of Cambridge Studies in Advanced Mathematics,
  Cambridge University Press, Cambridge, 1999.

\bibitem{Sc02}
{\sc W.~Schoutens}, {\em Meixner processes: Theory and applications in
  finance}, {EURANDOM} {R}eport 2002-004, EURANDOM, Eindhoven, 2002.

\bibitem{Ta11}
{\sc P.~Tankov}, {\em Pricing and hedging in exponential {L}\'evy models:
  review of recent results}, in Paris-{P}rinceton {L}ectures on {M}athematical
  {F}inance 2010, vol.~2003 of Lecture Notes in Math., Springer, Berlin, 2011,
  pp.~319--359.

\bibitem{Ya11}
{\sc S.~Yan}, {\em Jump risk, stock returns, and slope of implied volatility
  smile}, Journal of Financial Economics, 99 (2011), pp.~216--233.

\end{thebibliography}

\end{document}